\documentclass[twocolumn,aps,a4paper,superscriptaddress]{revtex4-2}
\usepackage{kpfonts}
\usepackage{amsmath}
\usepackage{amssymb}
\usepackage{times}
\usepackage{graphicx}
\usepackage[colorlinks,linkcolor=blue,citecolor=blue,urlcolor=blue,breaklinks]{hyperref}
\usepackage[normalem]{ulem}

\newcommand{\be}{\begin{equation}}
\newcommand{\ee}{\end{equation}}
\newcommand{\bea}{\begin{eqnarray}}
\newcommand{\eea}{\end{eqnarray}}

\def\a{\alpha}
\def\b{\beta}
\def\g{\gamma}
\def\G{\Gamma}
\def\d{\delta}
\def\D{\Delta}
\def\e{\epsilon}

\def\th{\theta}

\def\L{\Lambda}
\def\m{\mu}
\def\n{\nu}

\def\p{\pi}
\def\P{\Pi}
\def\r{\rho}
\def\s{\sigma}
\def\S{\Sigma}
\def\t{\tau}

\def\vf{\varphi}

\def\w{\omega}
\def\W{\Omega}
\def\q{\psi}
\def\Q{\Psi}

\def\ble{{\mathbf e}}

\def\blk{{\mathbf k}}

\def\bln{{\mathbf n}}

\def\blq{{\mathbf q}}
\def\blr{{\mathbf r}}

\def\blx{{\mathbf x}}

\def\blP{{\mathbf P}}

\def\blR{{\mathbf R}}

\def\blU{{\mathbf U}}

\def\callH{\mbox{$\mathcal{H}$}}

\def\de{\partial}
\def\iif{\infty}
\def\bra{\langle}
\def\ket{\rangle}

\def\1op{\hat{\mathbbm{1}}}
\def\nn{\nonumber}

\def\bz{\mathbf 0}

\begin{document}

\title{
The First Principles Equation for Coherent Phonons: Dynamics and 
Polaron distortions}

\author{Gianluca Stefanucci}%
 \email{gianluca.stefanucci@roma2.infn.it}
\affiliation{Dipartimento di Fisica, Universit{\`a} di Roma Tor Vergata, Via della Ricerca Scientifica 1,
00133 Rome, Italy}
\affiliation{INFN, Sezione di Roma Tor Vergata, Via della Ricerca Scientifica 1, 00133 Rome, Italy}

\author{Enrico Perfetto}%
\affiliation{Dipartimento di Fisica, Universit{\`a} di Roma Tor Vergata, Via della Ricerca Scientifica 1,
00133 Rome, Italy}
\affiliation{INFN, Sezione di Roma Tor Vergata, Via della Ricerca Scientifica 1, 00133 Rome, Italy}

\begin{abstract}
This paper addresses the first principles description of coherent phonons 
in systems subjected to optical excitations and/or doping. 
We reformulate the 
first-principles Ehrenfest equation (fpEE) [Phys. Rev. X {\bf 13}, 031026 (2023)]
in terms of Born-Oppenheimer  phonon frequencies and dynamical Born 
effective charges. We demonstrate that nonadiabatic 
effects renormalize the Born-Oppenheimer frequencies and introduce a damping term 
responsible for the finite lifetime of coherent phonons. Notably, 
both the frequency
renormalization and the lifetime are identical to those of quantum 
phonons. Furthermore, we show that electrons exert a force driven 
by an unconventional dynamically screened electron-phonon coupling. 
This coupling is smaller than the bare one even in the adiabatic limit, 
highlighting the need to revise current models. 
The fpEE is also used to develop a first-principles polaron theory 
that describes lattice distortions induced by doping.         
\end{abstract}

\maketitle

\section{Introduction}
The impressive progress in laser technology have paved the way to the 
optical control of lattice dynamics. Notable examples are 
light-driven 
metal-to-insulator~\cite{liu_terahertz_2012,nicholson_beyond_2018,horstmann_coherent_2020} 
and insulator-to-metal~\cite{cavalleri_femtosecond_2001,chavez_band_2018} 
transitions, charge density 
waves~\cite{tomeljak_dynamics_2009,wall_atomistic_2012,huber_coherent_2014}, structural phase 
transitions~\cite{cavalleri_femtosecond_2001,rini_control_2007,frigge_optically_2017,zhang_light-induced_2019,torre_colloquium_2021} 
and enhanced 
superconductivity~\cite{fausti_light-induced_2011,mitrano_possible_2016,buzzi_photomolecular_2020},
to mention a few. This rich phenomenology is always associated with 
distortions of the underlying equilibrium lattice, commonly referred to as {\em 
coherent phonons}~\cite{zhai_coherent_2024}. Due to electron-nuclear interactions the 
excitation of coherent phonons is accompanied with a reshaping of both the 
phononic and electronic band structures, along with a redistribution 
of phononic and electronic populations. 
Ultrafast resonant optical excitations of duration shorter than optical 
phonon frequencies (displacive 
excitations)~\cite{zeiger_theory_1992,kuznetsov_theory_1994,ishioka2009coherent,lakehal_microscopic_2019,fischer_microscopic_2025}
generate time-dependent 
coherent phonons that oscillate on timescales 
up to the picosecond range~\cite{perfetto_real-time_2023}. These oscillations can be exploited to 
design phonon-driven Floquet 
matter~\cite{hubener_phonon_2018,shin_phonon-driven_2018,wang_coherent-phonon_2023} or to
induce time-dependent shifts (in the sub meV range) of exciton 
absorption 
lines~\cite{jeong_coherent_2016,trovatello_strongly_2020,li_single-layer_2021,mor_photoinduced_2021,perfetto_theory_2024}.

In addition to optically driven sources, coherent phonons can also be 
generated through doping. In fact, a polaron is simply an added electron 
that is partially or entirely trapped by the lattice 
distortion~\cite{franchini_polarons_2021,alexandrov_advances_2010}. In 
this context, we may speak of steady-state coherent phonons.
Density Functional Perturbation Theory calculations of  
lattice distortions~\cite{sio_ab-initio_2019,sio_polarons_2019} 
rely on the Born-Oppenheimer (BO) approximation and 
do not account for electron-phonon ($e$-$ph$)
correlations. These limitations have recently
been  overcome using the many-body Green's function 
approach~\cite{lafuente_unified_2022,lafuente_ab-initio_2022}.

Although high performance computing have foster studies of 
coherent phonons from a microscopic 
perspective~\cite{caruso_quantum_2023,emeis_coherent_2024}, 
the underlying microscopic 
$e$-$ph$ Hamiltonian is not derived from first principles. 
A glaring evidence of this fact is the appearance of the BO 
phonon frequencies in the equation governing the dynamics of coherent 
phonons, while the first principles $e$-$ph$ Hamiltonian only 
contains bare phonon frequencies~\cite{marini_many-body_2015,stefanucci_in-and-out_2023}.

Coherent phonons satisfy the first-principles Ehrenfest equation 
(fpEE) derived in Ref.~\cite{stefanucci_in-and-out_2023}. As the fpEE does 
not contain dressed phonon frequencies, assessing the solidity of 
previous studies and developing a systematic approach to improve 
earlier methods is not a straightforward task. 
In this work, we reformulate the fpEE in terms of BO
phonon frequencies and examine the underlying simplifications 
that must be applied to recover the equations currently in use. 
Our first main conclusion is that the current equation of motion for 
coherent phonons must be revised, as it does not arise from the fpEE, 
even ignoring nonadiabatic effects. The main difference is 
that the $e$-$ph$ coupling should be replaced by an unconventional  
type of screened coupling, which, to the best of our knowledge, 
has not been previously discussed. The inclusion of nonadiabatic 
effects renormalizes the BO phonon frequencies, gives a 
finite lifetime to the coherent phonons, and leads to the dynamical 
generalization of the Born effective charge tensor~\cite{bistoni_giant_2019,binci_first-principles_2021,wang_dynamical_2022,dreyer_nonadiabatic_2022}. Notably, 
the frequency renormalization and lifetime are the same as 
for quantum phonons.
Our second main conclusion is that 
the first principles polaron approach developed in 
Refs.~\cite{lafuente_unified_2022,lafuente_ab-initio_2022} can be 
justified with reasonable assumptions.

The paper is organized as follows. In Section~\ref{bvdwsec} we 
revisit the concept of bare and BO phonon frequencies, and prove that 
they are both positive. The comparison between the first-principles 
$e$-$ph$ Hamiltonian and typical model Hamiltonians is presented in 
Section~\ref{fpvmsec}. The fpEE theory of coherent phonons is 
presented in Section~\ref{fpeeseccp}. In Section~\ref{qpappsec} we 
develop the quasi-phonon approximation and compare the resulting 
equation of motion with that emerging from model Hamiltonians 
or phenomenological approaches. 
In Section~\ref{polsec} we 
develop a quasiparticle theory for polaron distortions. We summarize 
the main findings in Section~\ref{sumconcsec}.

\section{Bare and BO phonon frequencies}
\label{bvdwsec}

In this section we briefly revisit the connection between 
bare and BO phonon frequencies. Let us consider a crystal
with $N_{n}$ nuclei, and denote by $\hat{\blR}_{i}$ and 
$\hat{\blP}_{i}$  the position and momentum operators 
of nucleus $i$, and by $\hat{\q}^{\dag}(\blx=\blr\s)$ the  
field operator for electrons in position $\blr$ and spin $\s$. 
The starting point is the 
Hamiltonian for electrons and nuclei
\begin{align}
\hat{H}&=\sum_{i=1}^{N_{n}}\frac{\hat{P}_{i}^{2}}{2M_{i}}+
\!\int \!d\blx \,
\hat{\q}^{\dag}(\blx)
\Big[-\frac{\nabla^{2}}{2}+V(\blr,\hat{\blR})
\Big]\hat{\q}(\blx)
\nn\\
&+E_{n-n}(\hat{\blR})+\hat{H}_{e-e},
\label{elnufull}
\end{align}
where $\hat{P}_{i}^{2}=\hat{\blP}_{i}\cdot \hat{\blP}_{i}$ is the 
squared momentum operator  of nucleus $i$, 
$V(\blr,\hat{\blR})= -\sum_{i=1}^{N_{n}}Z_{i}
v(\blr,\hat{\blR}_{i})$ is the electron-nuclear interaction potential, 
$E_{n-n}(\hat{\blR})= \frac{1}{2}\sum_{i\neq j}^{N_{n}}Z_{i}Z_{j}
v(\hat{\blR}_{i},\hat{\blR}_{j})$ is the nuclear-nuclear interaction 
Hamiltonian, and 
$\hat{H}_{e-e}$ is the electron-electron interaction Hamiltonian.
The Born-Oppenheimer (BO) Hamiltonian is defined as 
$\hat{H}_{BO}=\lim_{M_{i}\to\iif}\hat{H}$.
Since the nuclear position operators commute with $\hat{H}_{BO}$, all 
eigenstates of this operator have the form $|\blR\ket|\Q\ket$, where 
$|\Q\ket$ is a purely electronic ket. In particular we have
\begin{align}
\hat{H}_{BO}|\blR\ket|\Q\ket=|\blR\ket\hat{H}_{BO}(\blR)|\Q\ket
\end{align}
where $\hat{H}_{BO}(\blR)$ is a purely electronic operator obtained 
by replacing $\hat{\blR}\to\blR$ in $\hat{H}_{BO}$.
The following analysis is based on the  
working hypothesis that the equilibrium nuclear coordinates of
the BO Hamiltonian $\hat{H}_{BO}$ (infinite nuclear masses) give an 
excellent approximation to the true nuclear coordinates. This 
(often implicit) hypothesis is assumed to be verified in most 
treatments.

The ground state $|\Q^{0}\ket|\blR^{0}\ket$ of energy $E^{0}$ of the 
BO Hamiltonian can be obtained through 
two distinct double minimization procedures.
Let us consider the functional
\begin{align}
E(\Q,\blR)=\bra\Q|\hat{H}_{BO}(\blR)|\Q\ket.
\end{align}
By definition this functional has an absolute minimum in $(\Q^{0}, 
\blR^{0})$, and equals the ground state energy at the absolute 
minimum: $E^{0}=E(\Q^{0},\blR^{0})$.
We define $\blR_{\Q}$ as the nuclear coordinates that fulfill
\begin{align}
E(\Q,\blR_{\Q})=\min_{\blR}E(\Q,\blR),
\end{align}
and $|\Q(\blR)\ket$ as the electronic ket that fulfills
\begin{align}
E(\Q(\blR),\blR)=\min_{\Q}E(\Q,\blR).
\end{align}
Therefore, the state $|\Q(\blR)\ket$ is the ground state $\hat{H}_{BO}(\blR)$.
We then have 
\begin{align}
E^{0}=\min_{\Q}E(\Q,\blR_{\Q})=\min_{\blR}E(\Q(\blR),\blR),
\end{align}
and 
\begin{subequations}
\begin{align}
\Q^{0}&=\Q(\blR^{0}),
\\
\blR^{0}&=\blR_{\Q^{0}}.
\end{align}
\end{subequations}
Using the above definitions, the following inequalities can be derived
\begin{align}
E^{0}_{BO}(\blR)\equiv 
E(\Q^{0},\blR)\geq E(\Q^{0},\blR_{\Q^{0}})=E^{0},
\label{ineq1}
\end{align}
\begin{align}
E_{BO}(\blR)\equiv E(\Q(\blR),\blR)\geq E(\Q^{0},\blR_{\Q^{0}})= 
E^{0},
\label{ineq2}
\end{align}
which are both saturated for $\blR=\blR^{0}$. The function 
$E_{BO}(\blR)$ is the well known BO energy.
As $E^{0}_{BO}(\blR)$ and $E_{BO}(\blR)$ have an absolute 
minimum in $\blR^{0}$, the Hessian calculated in 
$\blR^{0}$ is {\em positive definite} for both of them. 

Let $R_{i\a}$, $\a=x,y,z$, be the cartesian components of 
the vector $\blR_{i}$.
For the Hessian of $E^{0}_{BO}(\blR)$ we find 
\begin{align}
K^{0}_{i\a,j\b}&=\left.\frac{\de^{2}E^{0}_{BO}(\blR)}{\de R_{i\a}\de 
R_{j\b}}\right|_{\blR=\blR^{0}}
\nn\\&=
\int \!d\blx\,n^{0}_{BO}(\blx)g^{\rm DW}_{i\a,j\b}(\blr)+
\left.\frac{\de^{2} E_{n-n}(\blR)}{\de R_{i\a}\de 
R_{j\b}}\right|_{\blR=\blR^{0}},
\label{2derboener}
\end{align}
where 
\begin{align}
n^{0}_{BO}(\blx)=\bra\Q^{0}|\hat{n}(\blx)|\Q^{0}\ket
\label{bodens}
\end{align}
is the ground-state BO density, and 
\begin{align}
g^{\rm DW}_{i\a,j\b}(\blr)&\equiv \left.\frac{\de^{2} 
V(\blr,\blR)}{\de R_{i\a}\de R_{j\b}}
\right|_{\blR=\blR^{0}}=-\d_{ij}Z_{i}\frac{\de^{2}}{\de 
r_{\a}\de r_{\b}}v(\blr,\blR^{0}_{i})
\end{align}
is the so called Debye-Waller coupling~\cite{giustino_electron-phonon_2017}.
The Hessian $K^{0}$ is the elastic tensor, represented by the 
symbol  
$K$ in Ref.~\cite{stefanucci_in-and-out_2023}. In that work, we  
argued that $K^{0}$ might have negative eigenvalues. 
We have just shown that under the aforementioned  working hypothesis,
the inequality in Eq.~(\ref{ineq1}) ensures that $K^{0}$ is positive 
definite.

Similarly, we can evaluate the Hessian of $E_{BO}(\blR)$, known as 
the dynamical matrix, usually represented by  the symbol $D$ (the 
symbol $\callH$ was used in Ref.~\cite{stefanucci_in-and-out_2023}). 
In the following we use $D$ for the phonon Green's function,
and therefore we represent the 
dynamical matrix by the symbol $K$. 
Using the 
Hellmann-Feynman theorem, it is straightforward to show 
that~\cite{baroni_phonons_2001,stefanucci_in-and-out_2023}
\begin{align}
K_{i\a,j\b}
&=\left.\frac{\de^{2}E_{BO}(\blR)}{\de R_{i\a}\de 
R_{j\b}}\right|_{\blR=\blR^{0}}
\nn\\
&=K^{0}_{i\a,j\b}+\int d\blx d\blx' g_{i\a}(\blr)
\chi^{R}_{\rm clamp}(\blx,\blx';0)g_{j\b}(\blr'),
\label{omega}
\end{align}
where
\begin{align}
g_{i\a}(\blr)&\equiv \left.\frac{\de V(\blr,\blR)}{\de 
R_{i\a}}\right|_{\blR=\blR^{0}}=Z_{i}\frac{\de}{\de 
r_{\a}}v(\blr,\blR^{0}_{i})
\label{gia(r)}
\end{align}
is the {\em bare} $e$-$ph$ coupling and
$\chi^{\rm R}_{\rm clamp}$ is the density-density response 
function at clamped nuclei.

Let $(\w^{0}_{\n\blq})^{2}\geq 0$ and $(\w_{\n\blq})^{2}\geq 0$ be 
the eigenvalues of $K^{0}$ and $K$ respectively. They are all positive 
since $K^{0}$ and $K$ are positive definite. The energies 
$\w^{0}_{\n\blq}>0$ are said the {\em bare} phonon frequencies as they 
correspond to the frequencies of the normal 
modes of the nuclear lattice at {\em fixed} electrons. 
The energies $\w_{\n\blq}>0$ are said the BO frequencies and often provide 
a good estimate for the {\em dressed} phonon 
frequencies. 
Taking into account that $\chi^{\rm R}_{\rm clamp}(\blx,\blx';0)$ is 
{\em negative definite}, we expect that the bare 
frequencies are typically larger than the dressed ones. 
This is certainly true when all tensors in  Eq.~(\ref{omega}) have 
common eigenvectors, which is generally not the case.
Nonetheless, the expectation aligns with the intuitive picture 
that the eigenfrequencies of 
the lattice increase when the electrons are frozen, as freezing 
the electrons  makes the 
lattice more rigid.   

\section{First principles versus model Hamiltonians}
\label{fpvmsec}

We identify a generic nucleus $i$ with the vector (of integers) $\bln$ of 
the unit cell and with a label $s$, i.e., 
$\blR^{0}_{i=\bln s}=\blR^{0}_{\bln}+\blR^{0}_{s}$. 
Nuclei of type $s$ have the same mass and charge in all unit cells, 
$M_{i=\bln s}=M_{s}$ and $Z_{i=\bln s}=Z_{s}$.
We expand the displacement and momentum operators as
\begin{subequations}
\begin{align}
\hat{U}_{\bln s\a}=\frac{1}{\sqrt{M_{s}N}}
\sum_{\blq}e^{i\blq\cdot\bln}\,
\sum_{\n}
e^{\n}_{s\a}(\blq)
\;\hat{U}_{ \n\blq},
\label{displdecnm}
\end{align}
\begin{align}
\hat{P}_{\bln s\a}=\sqrt{\frac{M_{s}}{N}}
\sum_{\blq}e^{i\blq\cdot\bln}\,
\sum_{\n}
e^{\n}_{s\a}(\blq)
\;\hat{P}_{ \n\blq},
\label{momdecnm}
\end{align}
\label{momdisplexp}
\end{subequations}
where $N$ is the total number of cells, and 
the sum over $\blq=(q_{x},q_{y},q_{z})$ 
runs over vectors with components $q_{\a}$ 
in the range $(-\p,\p]$.
The vectors $\ble^{\n}(\blq)$, with components $e^{\n}_{s\a}(\blq)$, 
are referred to as {\em normal modes}. 
They form an orthonormal basis for each $\blq$, i.e., 
$\ble^{\n}(\blq)^{\dag}\cdot \ble^{\n'}(\blq)=\d_{\n\n'}$. 
It is straightforward to verify the commutation relation 
$[\hat{U}_{ \n\blq},\hat{P}^{\dag}_{\n'\blq'}]=i\d_{\n\n'}\d_{\blq,\blq'}$.
Due to the lattice periodicity, the bare and BO Hessians have the 
property 
\begin{align}
K^{0}_{\bln s\a,\bln's'\a'}&=K^{0}_{s\a,s'\a'}(\bln-\bln'),
\nn\\
K_{\bln s\a,\bln's'\a'}&=K_{s\a,s'\a'}(\bln-\bln').
\end{align}

The first principles $e$-$ph$ Hamiltonian follows from Eq.~(\ref{elnufull}) 
when expanding to second order in the nuclear displacements 
$\blU=\blR-\blR^{0}$ and density variations $\D n=n-n^{0}$. A 
detailed discussion on its derivation is given in 
Refs.~\cite{stefanucci_in-and-out_2023,svl-book_2025}. Here, we report the 
final result: 
\begin{align}
\hat{H}=\hat{H}_{0,e}+\hat{H}_{e-e}+\hat{H}_{e-ph}+\hat{H}_{0,ph}.
\label{el-phonham3}
\end{align}
In this equation, 
\begin{align}
\hat{H}_{0,e}=\!\int \!d\blx \,
\hat{\q}^{\dag}(\blx)
\Big[-\frac{\nabla^{2}}{2}+V(\blr,\blR^{0})
\Big]\hat{\q}(\blx)
\label{h0e}
\end{align}
is the free-electron part, $\hat{H}_{e-e}$ is
the electron-electron interaction Hamiltonian,
\begin{align}
\hat{H}_{e-ph}=\sum_{\blq\n}\int d\blx \;g_{\n-\blq}(\blr)
\hat{n}(\blx)
\hat{U}_{ \n\blq}
\label{el-phonintham}
\end{align}
is the $e$-$ph$ interaction Hamiltonian, 
$\hat{n}(\blx)=\hat{\q}^{\dag}(\blx)\hat{\q}(\blx)$ 
being the density operator, and 
\begin{align}
\hat{H}_{0,ph}&=
\frac{1}{2}\sum_{\blq \n}\hat{P}^{\dag}_{ \n\blq}
\hat{P}_{ \n\blq}
+\frac{1}{2}\sum_{\blq\n\n'}
\hat{U}^{\dag}_{ \n\blq}K^{0}_{\n\n'\blq}\,
\hat{U}_{ \n'\blq}
\nn\\&
-\sum_{\blq\n}\int d\blx \;g_{\n-\blq}(\blr)
n^{0}(\blx)
\;\hat{U}_{ \n\blq},
\label{H0phondexp}
\end{align}
is the bare phonon Hamiltonian. In Eqs.~(\ref{el-phonintham}) and 
(\ref{H0phondexp}) we have defined  
\begin{align}
K^{0}_{\n\n'\blq}&\equiv 
\sum_{\bln}e^{-i\blq\cdot\bln}
\sum_{s\a,s'\a'}
e^{\n\ast}_{s\a}(\blq)\;
\frac{K^{0}_{s\a,s'\b}(\bln)}{\sqrt{M_{s}M_{s'}}}
\;e^{\n'}_{s'\a'}(\blq)\nn\\&=K^{0\ast}_{\n\n'-\blq}
=K^{0\ast}_{\n'\n\blq},
\label{Kprop}
\end{align}
and
\begin{align}
g_{\n-\blq}(\blr)\equiv 
\sum_{\bln s \a}\frac{1}{\sqrt{M_{s}N}}
e^{i\blq\cdot\bln}\,e^{\n}_{s\a}(\blq)\,
g_{\bln s\a}(\blr)=g^{\ast}_{ \n\blq}(\blr).
\label{proelphoncoup}
\end{align}
Notice that neither the dressed phonon 
frequencies nor the dressed $e$-$ph$ coupling appear in 
$\hat{H}$. Also notice that $n^{0}(\blx)$ is the true (as opposed to 
BO) equilibrium density, which must be determined 
self-consistently~\cite{stefanucci_in-and-out_2023,svl-book_2025}.

For later reference, we compare the first-principles 
Hamiltonian with typical model Hamiltonians for $e$-$ph$ 
systems. The main and only difference is that $\hat{H}_{0,ph}\to \hat{H}_{0,ph}^{\rm model}$, 
where 
\begin{align}
\hat{H}_{0,ph}^{\rm model}=
\frac{1}{2}\sum_{\blq \n}\hat{P}^{\dag}_{ \n\blq}
\hat{P}_{ \n\blq}
+\frac{1}{2}\sum_{\blq\n\n'}
\hat{U}^{\dag}_{\n\blq}K_{\n\n'\blq}\,
\hat{U}_{ \n'\blq},
\label{H0phondexpmod}
\end{align}
and $K_{\n\n'\blq}$ is defined as in Eq.~(\ref{Kprop}) with 
$K^{0}\to K$. 
Thus, the Hamiltonian $\hat{H}_{0,ph}^{\rm model}$ contains the BO Hessian 
and omits any linear term in the displacements.
In the basis of the normal modes, the relation between $K$ and 
$K^{0}$, see Eq.~(\ref{omega}), reads
\begin{align}
K_{\n\n'\blq}=K^{0}_{\n\n'\blq}+
\int d\blx d\blx' g_{\n\blq}(\blr)
\chi^{\rm R}_{\rm clamp}(\blx,\blx';0)g^{\ast}_{\n'\blq}(\blr').
\label{omom0pi}
\end{align}
A convenient choice of normal modes is the eigenbasis of the 
BO Hessian. Henceforth, this is the basis we use, and therefore 
\begin{align}
K_{\n\n'\blq}=\d_{\n\n'}\w^{2}_{\n\blq}.
\label{bobasis}
\end{align}

We are also interested in driving the system out of equilibrium using 
external electromagnetic fields. For simplicity we consider a vector 
potential $A_{\a}(t)$, $\a=x,y,z$, that does not 
break the translational invariance.
Then, the Hamiltonian describing the 
light-matter coupling reads
\begin{align}
\hat{H}_{\rm ext}= \frac{1}{c}\sum_{\a}\left(\int d\blx \,
\hat{J}_{\a}(\blx t)-\sum_{\bln 
s}\frac{Z_{s}}{M_{s}}\hat{P}_{\bln s\a}\right)A_{\a}(t),
\label{hlm1}
\end{align}
where $\hat{J}_{\a}(\blx t)$ is the gauge invariant electronic 
current operator. The second term in Eq.~(\ref{hlm1}) accounts for 
the direct coupling 
between light and nuclei.
	
\section{First-principles Ehrenfest Equation for Coherent Phonons}
\label{fpeeseccp}

The fpEE for the average value of the nuclear 
displacement operators can easily be derived from the commutator between the 
Hamiltonian and the momentum operators. 
As the external field does not 
break the translational invariance, we consider only 
the $\blq=\bz$ component of the displacements. 
The fpEE  has been derived 
in Refs.~\cite{stefanucci_in-and-out_2023,svl-book_2025} and reads
\begin{align}
\frac{d^{2}U_{\n\bz}(t)}{dt^{2}}
&=-\sum_{\n'}K^{0}_{\n\n'\bz}U_{\n'\bz}(t)
-\int d\blx \;g_{\n\bz}(\blr)
\D n(\blx t)
\nn\\
&+\sum_{\a}Z_{\n\a}E_{\a}(t),
\label{ehr1}
\end{align}
where $E_{\a}(t)=-\frac{1}{c}\frac{dA_{\a}(t)}{dt}$ is the electric field,
\begin{align}
Z_{\n\a}\equiv\sum_{s}\sqrt{\frac{N}{M_{s}}}e^{\n}_{s\a}(\bz)Z_{s},
\label{bBOcharge}
\end{align}
and
\begin{align}
\D n(\blx t)\equiv n(\blx t)-n^{0}(\blx)
\end{align}
is the density fluctuation.
Equation (\ref{ehr1}) should be compared with the equation of motion 
generated by the model Hamiltonian in Eq.~(\ref{H0phondexpmod}). Denoting by $U^{\rm model}$ and 
$n^{\rm model}$ the 
resulting time-dependent displacement and density, we find
\begin{align}
\frac{d^{2}U^{\rm model}_{\n\bz}(t)}{dt^{2}}
&=-\w^{2}_{\n\bz}U^{\rm model}_{\n\bz}(t)
-\int d\blx \;g_{\n\bz}(\blr)
n^{\rm model}(\blx,t)
\nn\\
&+\sum_{\a}Z_{\n\a}E_{\a}(t).
\label{ehr1mod}
\end{align}
One notable issue with this equation is that the equilibrium 
solution is not $U^{\rm model}_{\n\bz}=0$, since the full density, 
rather than the density fluctuation, appears on the right-hand side.   
A second crucial difference with the fpEE is that the displacement on 
the r.h.s. is multiplied by the BO Hessian rather than the bare 
Hessian.
It is the purpose of 
this section to manipulate Eq.~(\ref{ehr1}) into a form as close as 
possible  to Eq.~(\ref{ehr1mod}), allowing us to appreciate 
the differences between them. 

We write the full density fluctuation as
\begin{align}
\D n(\blx t)=\D n^{(1)}(\blx t)	+\D n^{(r)}(\blx t),
\end{align}
where $\D n^{(1)}$ is linear in the external field and  
the remainder $\D n^{(r)}$ accounts for nonlinear contributions. 
As we see in Section~\ref{semisubsec}, in semiconductors 
the linear change is proportional to the polarization whereas 
the remainder is dominated by populations.
The linear contribution is given by~\cite{svl-book_2025}
\begin{align}
\D n^{(1)}(\blx t)=\frac{1}{c}\sum_{\a}\int d(\blx't')\chi^{R}_{0\a}(\blx t,\blx't')
A_{\a}(t'),
\label{dn1.1}
\end{align}
where $\chi^{R}_{0\a}$, $\a=x,y,z$, is the (retarded) density-current response 
function of the system in equilibrium. 
As discussed in 
Refs.~\cite{stefanucci_semiconductor_2024,stefanucci_excitonic_2025,stefanucci_exact_2025}, this 
response function satisfies the Dyson equation (in frequency space)
\begin{widetext}
\begin{align}
\chi^{R}_{0\a}(\blx,\blx';\w)&=
\tilde{\chi}^{R}_{0\a}(\blx,\blx';\w)
+\sum_{\blq\bar{\n}\bar{\n}'}
\int d\bar{\blx}d\bar{\blx}'
\tilde{\chi}^{R}_{00}(\blx,\bar{\blx};\w)
g^{\ast}_{\bar{\n}\blq}(\bar{\blr})D^{0,R}_{\bar{\n}\bar{\n}'\blq}(\w)
g_{\bar{\n}'\blq}(\bar{\blr}')\chi^{R}_{0\a}(\bar{\blx}',\blx';\w),
\label{chidyson}
\end{align}
where $\tilde{\chi}$ is the {\em phonon-irreducible} response function and 
$D^{0}$ is the {\em bare} phonon propagator: 
$D^{0,R}_{\n\n'\blq}(\w)=\left[
\frac{1}{(\w+i\eta)^{2}-K^{0}_{\blq}}\right]_{\n\n'}$.
We observe that the diagrammatic expansion of the phonon-irreducible 
density-density response function $\tilde{\chi}^{R}_{00}$ in Eq.~(\ref{chidyson}) contains 
electron-electron and $e$-$ph$ vertices. Discarding all diagrams 
with $e$-$ph$ vertices yields the density-density response function 
at clamped nuclei $\chi^{R}_{\rm clamp}$ appearing 
in Eq.~(\ref{omega}).

Using Eqs.~(\ref{dn1.1}) and (\ref{chidyson}) 
we show in Appendix~\ref{fpEEapp} that the fpEE can 
be rewritten as 
\begin{align}
\frac{d^{2}U_{\n\bz}(t)}{dt^{2}}
&=-\sum_{\n'}\Big[K^{0}_{\n\n'\bz}U_{\n'\bz}(t)
-\int dt'\,
\P^{R}_{\n\n'\bz}(t,t')U_{\n'\bz}(t')\Big]
-\int d(\blx't')\tilde{g}^{s}_{\n\bz}(\blr';t,t')
\D n^{(r)}(\blx't')+F_{\n}(t).
\label{ehr3}
\end{align}
\end{widetext}
This is first main result of our work. It is expressed in terms of three key 
quantities.

The first quantity is the equilibrium phonon self-energy
\begin{align}
\P^{R}_{\n\n'\blq}(t,t')=\int d\blx d\blx'\;
g_{\n\blq}(\blr)
\tilde{\chi}^{R}_{00}(\blx t,\blx't')
g^{\ast}_{\n'\blq}(\blr')
\label{phonself}
\end{align}	
evaluated in $\blq=\bz$, which is correctly expressed in terms of the 
phonon-irreducible response function $\tilde{\chi}$, 
not the full response function $\chi$~\cite{stefanucci_exact_2025}. 

The second quantity is the screened external force  acting on the nuclei [compare with last term 
in Eq.~(\ref{ehr1})]
\begin{align}
F_{\n}(t)&\equiv\sum_{\a}\int\frac{d\w}{2\p}e^{-i\w t}\frac{i\w}{c}Z^{\rm 
eff}_{\n\a}(\w)A_{\a}(\w),
\label{scrforce}
\end{align}
where $Z^{\rm eff}_{\n\a}(\w)$ is the  {\em dynamical} 
(or nonadiabatic) Born effective charge 
tensor~\cite{bistoni_giant_2019,binci_first-principles_2021,wang_dynamical_2022,dreyer_nonadiabatic_2022}
in the basis of BO normal modes [compare with Eq.~(\ref{bBOcharge})]:
\begin{align}
Z^{\rm eff}_{\n\a}(\w)&=
\sum_{\bln s \b}\frac{e^{\n}_{s\b}(\bz)}{\sqrt{M_{s}N}}
\nn\\
&\times\left[Z_{s}\d_{\a\b}+\frac{i}{\w}\int d\blx d\blx' 
g_{\bln s\b}(\blr)
\tilde{\chi}^{R}_{0\a}(\blx,\blx';\w)\right].
\end{align}
For optical fields in the visible range, this force has no 
impact on the nuclear coordinates as it oscillates too fast to set 
the  lattice in motion. In contrast, for infrared fields, 
the $F_{\n}$ 
force provides a direct mechanism for exciting coherent phonons. 

The third quantity is the unconventional dynamically screened $e$-$ph$ coupling
\begin{align}
\tilde{g}^{s}_{\n\bz}(\blr;t,t')&\equiv
\sum_{\bar{\n}'}\int\frac{d\w}{2\p}e^{-i\w(t-t')}
\nn\\
&\times \big[\d_{\n\bar{\n}'}-\sum_{\bar{\n}}
\P^{R}_{\n\bar{\n}\bz}(\w)
D^{0,R}_{\bar{\n}\bar{\n}'\bz}(\w)\big]
 g_{\bar{\n}'\bz}(\blr),
\label{unconscreen}
\end{align}
which is
distinct from the screened $e$-$ph$ coupling,
$g^{s}=(1+v\tilde{\chi}^{R}_{00})g$, 
governing the phonon 
linewidths~\cite{allen_neutron_1972,grimvall_the-electron-phonon_1981,stefanucci_exact_2025}.
Since $\tilde{g}^{s}$ multiplies the nonlinear change in the 
electronic density, the fpEE is not closed with respect to $U_{\n\bz}$ and 
must therefore be coupled to the Kadanoff-Baym equations 
for the electronic and phononic Green's 
functions~\cite{stefanucci_in-and-out_2023}.
Alternatively, simplified versions of these equations can be used, 
such as the semiconductor electron-phonon 
equations~\cite{stefanucci_semiconductor_2024}, semiconductor 
Bloch equations~\cite{haug2009quantum, kira_many-body_2006},   
or Boltzmann 
equations~\cite{ziman_electrons_1960,kb-book,danielewicz_quantum_1984,sadasivam_theory_2017,ponce_toward_2018,caruso_ultrafast_2022}.
This strategy has been explored in Refs.~\cite{emeis_coherent_2024,perfetto_theory_2024}.
For low intensities, we can expand $\D 
n^{(r)}$ to second order in the external field: $\D 
n^{(r)}=\sum_{\a\b}\chi^{(2)}_{0\a\b}A_{\a}A_{\b}$, where $\chi^{(2)}$
is the second-order response function. 
In this way, the fpEE becomes a close equation  
for the light-induced generation of coherent phonons 
of both impulsive and displacive 
character~\cite{garrett_coherent_1996,merlin_generating_1997,stevens_coherent_2002}. Our first-principles 
derivation shows that the analysis of 
Refs.~\cite{garrett_coherent_1996,merlin_generating_1997,stevens_coherent_2002} remains correct provided that the 
$e$-$ph$ coupling is replaced by a time-local approximation of $\tilde{g}^{s}$.

The fpEE in Eq.~(\ref{ehr3}) is the closest we can make Eq.~(\ref{ehr1}) 
resemble Eq.~(\ref{ehr1mod}).
We emphasize that Eqs.~(\ref{ehr1}) and (\ref{ehr3}) are both  exact 
results. They are two equivalent ways of rewriting the fpEE.
The advantage of Eq.~(\ref{ehr3}) lies in its ease of manipulation and simplification
when physically insightful approximations are applied.
In the next section we discuss the fpEE in the quasi-phonon 
approximation,
along with a few interesting consequences.

\section{Quasi-phonon approximation}
\label{qpappsec}

We separate the phonon self-energy $\P^{R}(t,t')$
into a Drude-like contribution and a dynamical contribution:
\begin{align}
\P^{R}_{\n\n'\bz}(t,t')=\P^{R}_{\n\n'\bz}(0)\d(t,t')+\D\P^{R}_{\n\n'\bz}(t,t'),
\label{anasplit}
\end{align}
where $\P^{R}_{\n\n'\bz}(0)$ is the Fourier transform of the phonon 
self-energy at {\em clamped nuclei} 
evaluated at zero frequency. 
In frequency space Eq.~(\ref{anasplit}) reads
\begin{align}
\P^{R}_{\n\n'\bz}(\w)=\P^{R}_{\n\n'\bz}(0)+\D\P^{R}_{\n\n'\bz}(\w).
\end{align}
The first term is commonly referred to as the adiabatic contribution, 
whereas the second term is referred to as the nonadiabatic 
contribution~\cite{giustino_electron-phonon_2017}. Following 
Refs.~\cite{giustino_electron-phonon_2017,stefanucci_in-and-out_2023} 
we further approximate the nonadiabatic contribution by discarding 
the off-diagonal entries:
\begin{align}
\D\P^{R}_{\n\n'\bz}(\w)\simeq 
\d_{\n\n'}\big[\L_{\n\bz}(\w)-\frac{i}{2}\G_{\n\bz}(\w)\big],
\label{deltapisplit}
\end{align}
where $\L_{\n\bz}(\w)$ and $\G_{\n\bz}(\w)\geq 0$ are real functions.

With these approximations the first integral in Eq.~(\ref{ehr3}) 
becomes
\begin{align}
\sum_{\n'}\int dt'\,
\P^{R}_{\n\n'\bz}(t,t')U_{\n'\bz}(t')&=
\sum_{\n'}\P^{R}_{\n\n'\bz}(0)U_{\n'\bz}(t)
\nn\\
&+\int dt'\,
\D\P^{R}_{\n\n\bz}(t,t')U_{\n\bz}(t').
\label{intstep}
\end{align}
To evaluate the second term in Eq.~(\ref{intstep})
we assume that the solution of 
Eq.~(\ref{ehr3})  is well approximated by
\begin{align}
U_{\n\bz}(t)=u_{\n\bz}(t)e^{i\W_{\n\bz}t}+
u^{\ast}_{\n\bz}(t)e^{-i\W_{\n\bz}t},
\label{ansatzunu}
\end{align}
where $u_{\n\bz}(t)$ is a slowly varying function of time and 
$\W_{\n\bz}\geq 0$ is a frequency to be determined. 
Ignoring the dependence 
on time of the function $u_{\n\bz}(t)$ it is straightforward to obtain
\begin{align}
\int \!d\bar{t}\,
\D\P^{R}_{\n\n\bz}(t,\bar{t})U_{\n\bz}(\bar{t})=
\big[\L_{\n\bz}(\W_{\n\bz})+\frac{\G_{\n\bz}(\W_{\n\bz})}{2\W_{\n\bz}}
\frac{d}{dt}\big]U_{\n\bz}(t),
\end{align}
where we use the symmetry properties $\L_{\n\bz}(\w)=\L_{\n\bz}(-\w)$
and $\G_{\n\bz}(\w)=-\G_{\n\bz}(-\w)$ -- see Appendix C of 
Ref.~\cite{stefanucci_in-and-out_2023}.

Next we examine the second integral in Eq.~(\ref{ehr3}). For 
general external drivings, and thus for general time-dependent 
functions $\D n^{(r)}(\blx' t')$, this term can be hardly 
approximated. However, for ultrafast optical drivings (duration 
shorter than a typical optical phonon period) $\D 
n^{(r)}(\blx' t')$ changes rapidly during the drive, and 
can be approximated as a slow function of time 
thereafter. For lasers impinging the system at time, say, $t=0$, we
approximate $\D n^{(r)}(\blx' t')\simeq \th(t')
\D n^{(r)}(\blx')$, and hence in Fourier space
$\D n^{(r)}(\blx' \w)\simeq i\D n^{(r)}(\blx')/(\w+i\eta)$.
We then have
\begin{align}
\int\frac{d\w}{2\p}e^{-i\w t}\big[\d_{\n\bar{\n}'}-\sum_{\bar{\n}}
\P^{R}_{\n\bar{\n}\bz}(\w)
D^{0,R}_{\bar{\n}\bar{\n}'\bz}(\w)\big]\frac{i\D n^{(r)}(\blx' )}{\w+i\eta}
\nn\\
\simeq \big[\d_{\n\bar{\n}'}+\sum_{\bar{\n}}\P^{R}_{\n\bar{\n}\bz}(0)
[K^{0}_{\bz}]^{-1}_{\bar{\n}\bar{\n}'}\big]\D n^{(r)}(\blx' t),
\label{secondpiece}
\end{align}
where in the last equality we neglect the contributions from the 
poles of $\P^{R}_{\n\bar{\n}\bz}(\w)
D^{0,R}_{\bar{\n}\bar{\n}'\bz}(\w)$, and use that
$D^{0,R}_{\n\n'\blq}(0)=-[K^{0}_{\blq}]^{-1}_{\n\n'}$. 
We observe that Eq.~(\ref{secondpiece}) can alternatively be derived 
by implementing the Markov approximation $\D n^{(r)}(\blx' t')\simeq 
\D n^{(r)}(\blx' t)$ under the integral sign of Eq.~(\ref{ehr3}).

Using the above approximations -- henceforth collectively referred to 
as the {\em quasi-phonon 
approximation} -- the fpEE in Eq.~(\ref{ehr3})
reduces to an ordinary differential equation for the displacements:
\begin{align}
\frac{d^{2}U_{\n\bz}(t)}{dt^{2}}
&=-\big[\w^{2}_{\n\bz}+\L_{\n\bz}(\W_{\n\bz})\big]
U_{\n\bz}(t)-\frac{\G_{\n\bz}(\W_{\n\bz})}{2\W_{\n\bz}}
\frac{dU_{\n\bz}(t)}{dt}
\nn\\
&-\int d\blx'
\,\tilde{g}^{s}_{\n\bz}(\blr')\D n^{(r)}(\blx' t)+F_{\n}(t),
\label{ehr3stat}
\end{align}
where we take into account Eqs.~(\ref{omom0pi}) and (\ref{bobasis}).
Indeed, the phonon self-energy $\P^{R}_{\bz}(0)$ is, by definition, 
evaluated at clamped nuclei, and therefore
coincides with the second term in 
Eq.~(\ref{omom0pi}) -- recall that at clamped nuclei
$\tilde{\chi}_{00}=\chi_{\rm clamp}$, see discussion below 
Eq.~(\ref{chidyson}).
Consequently, the sum 
$K^{0}_{\bz}+\P^{R}_{\bz}(0)$ is the BO Hessian $K_{\bz}$.
As we are working in the eigenbasis of the BO Hessian, 
$K_{\bz}$ is diagonal, see Eq.~(\ref{bobasis}). 
In deriving Eq.~(\ref{ehr3stat}) we also use that 
the quasi-phonon approximation makes the unconventional screened 
$e$-$ph$ coupling, see 
Eq.~(\ref{unconscreen}), time local. Indeed, from 
Eq.~(\ref{secondpiece}) we infer that
$\tilde{g}^{s}_{\n\bz}(\blr;t,t')=\d(t,t')\tilde{g}^{s}_{\n\bz}(\blr)$
with
\begin{equation}
\tilde{g}^{s}_{\n\bz}(\blr)=
\sum_{\n'}
\w^{2}_{\n\bz}[K^{0}_{\bz}]^{-1}_{\n\n'}g_{\n'\bz}(\blr).
\label{qpgamma}
\end{equation}

The result in Eq.~(\ref{ehr3stat}) indicates how to choose the frequency 
$\W_{\n\bz}$. It is determined from the solution of 
\begin{align}
\W_{\n\bz}=\sqrt{\w^{2}_{\n\bz}+\L_{\n\bz}(\W_{\n\bz})}\simeq
\w_{\n\bz}+\frac{\L_{\n\bz}(\w_{\n\bz})}{2\w_{\n\bz}},
\end{align}
i. e., $\W_{\n\bz}$ is the same as the renormalized  
frequency of quantum phonons~\cite{giustino_electron-phonon_2017,stefanucci_in-and-out_2023}. 
Interestingly, the nonadiabatic contribution is also responsible for 
a damping term, with damping time
\begin{align}
\t_{\n\bz}=\frac{2\W_{\n\bz}}{\G_{\n\bz}(\W_{\n\bz})},
\end{align}
which is exactly the same as the lifetime of quantum 
phonons~\cite{giustino_electron-phonon_2017,stefanucci_in-and-out_2023}.
In terms of the quantities just introduced, the quasi-phonon 
approximation of the fpEE,  Eq.~(\ref{ehr3stat}),
reads
\begin{align}
\frac{d^{2}U_{\n\bz}(t)}{dt^{2}}
&=-\W_{\n\bz}^{2}
U_{\n\bz}(t)-\frac{1}{\t_{\n\bz}}
\frac{dU_{\n\bz}(t)}{dt}
\nn\\&-\int d\blx'
\,\tilde{g}^{s}_{\n\bz}(\blr')\D n^{(r)}(\blx' t)+F_{\n}(t).
\label{ehrqp}
\end{align}

We emphasize that the model equation of motion Eq.~(\ref{ehr1mod}) is not 
recovered even discarding nonadiabatic effects -- responsible for the 
renormalization of the BO frequency and the damping term. The 
 differences are: (1)  the term $g_{\n\bz}$ is replaced by 
 $\tilde{g}^{s}_{\n\bz}$,  (2) 
instead of the full density $n$ we have the {\em nonlinear} change in 
the density $\D n^{(r)}$, and (3) the bare force 
$\sum_{\a}Z_{\n\a}E_{\a}$ is replaced by the screened force $F_{\n}$. 

It is also instructive to compare the fpEE 
with a phenomenological equation of 
motion that is 
widely used in the 
literature~\cite{bron_picosecond_1986,merlin_generating_1997,dekorsy_coherent_2000,juraschek_sum-frequency_2018,zhai_coherent_2024} 
\begin{align}
\frac{d^{2}U^{\rm phen}_{\n\bz}(t)}{dt^{2}}
&=-\W_{\n\bz}^{2}
U_{\n\bz}(t)-\g^{\rm phen}_{\n}
\frac{dU^{\rm phen}_{\n\bz}(t)}{dt}
\nn\\&+\sum_{\a\b}R_{\n\a\b}E_{\a}(t)E_{\b}(t)
+\sum_{\a}Z^{\rm eff}_{\n\a}E_{\a}(t),
\label{ehrqpphen}
\end{align}
where $R_{\n\a\b}$ is the Raman tensor and $Z^{\rm eff}_{\n\a}=Z^{\rm 
eff}_{\n\a}(0)$ is 
the {\em static} Born effective charge tensor.
The quasi-phonon approximation of the fpEE, Eq.~(\ref{ehrqp}), 
reduces to the phenomenological equation provided that 
$1/\g^{\rm phen}_{\n}$ is identified with the lifetime of the quantum phonon, 
$Z^{\rm eff}_{\n\a}(\w)\simeq Z^{\rm eff}_{\n\a}(0)$ for all $\w$, 
and $\D n^{(r)}(\blx' t)$ is expanded to second order in the external 
field as outlined in 
Refs.~\cite{garrett_coherent_1996,merlin_generating_1997,stevens_coherent_2002} --
this also implies that the Raman 
tensor, to be used for  coherent phonon dynamics, must be 
redefined in terms of the unconventional screened coupling 
$\tilde{g}^{s}$.
A straightforward improvement of the phenomenological equation consists 
in replacing $Z^{\rm eff}_{\n\a}$ with $Z^{\rm 
eff}_{\n\a}(\w_{0})$, where $\w_{0}$ is the frequency of the 
electromagnetic field.

Below we apply Eq.~(\ref{ehrqp}) to
semiconductors and  the jellium model.

\subsection{Semiconductors} 
\label{semisubsec}

In the basis of electronic 
Bloch functions $|n\blk\ket$, with $n$ the band index and $\blk$ the 
quasi-momentum, we can rewrite Eq.~(\ref{ehrqp}) as (omitting the 
explicit dependence on time)
\begin{align}
\frac{d^{2}U_{\n\bz}}{dt^{2}}
=-\W_{\n\bz}^{2}
U_{\n\bz}-\frac{1}{\t_{\n\bz}}
\frac{dU_{\n\bz}}{dt}
-\sum_{nm\blk}
\tilde{g}^{s,nm}_{\n\bz}(\blk)\D \r^{(r)}_{mn\blk},
\label{ehr3statclamp2}
\end{align}
where
\begin{align}
\tilde{g}^{s,nm}_{\n\blq}(\blk)
=\bra n\blk|\tilde{g}^{s}_{\n\blq}(\hat{\blr})|m\blk+\blq\ket,
\label{ginkspace}
\end{align}
and $\D \r^{(r)}$ is the nonlinear change in the density matrix.
For simplicity we here ignore the screened force $F_{\n}$.

In semiconductors with filled valence bands and 
empty conduction bands the electronic occupations do not 
change to first order in the external fields, i.e.,
\begin{align}
\D \r^{(r)}_{nn\blk}(t)=\D 
\r_{nn\blk}(t)=(f_{n\blk}(t)-f_{n\blk}^{\rm eq}),
\end{align}
with $f_{n\blk}(t)$ the carrier occupation at time $t$ and 
$f_{n\blk}^{\rm eq}$ the carrier occupation in equilibrium.
Assuming that  
the dominant contribution to the nonlinear 
change in the density matrix is given by the diagonal entries, 
we may approximate $\D \r^{(r)}_{mn\blk}\simeq \d_{mn}\D \r_{mm\blk}$ in 
Eq.~(\ref{ehr3statclamp2}). In this way the fpEE gets
closer to the model Ehrenfest
equation of motion, Eq.~(\ref{ehr1mod}), since the full density matrix 
(not just the remainder) appears. This analysis establishes the 
conditions under which the model equation is applicable~\cite{perfetto_theory_2024}.

Further assuming that $K$ and $K^{0}$ are diagonal in the same 
basis, the unconventional screened $e$-$ph$ coupling 
of Eq.~(\ref{qpgamma}) simplifies to
\begin{align}
\tilde{g}^{s,nm}_{\n\bz}(\blk)=
\left(\frac{\w_{\n\bz}}{\w^{0}_{\n\bz}}\right)^{2}
g^{nm}_{\n\bz}(\blk).
\end{align}
As we already observed $\w_{\n\bz}\leq 
\w^{0}_{\n\bz}$ since $\P^{R}_{\bz}(0)$ is negative definite. 
Therefore, the magnitude of the unconventional screened coupling 
is always smaller than the magnitude of the bare $e$-$ph$ coupling.

\subsection{Jellium} 

The simplest metallic system to investigate $e$-$ph$ interactions is 
the homogeneous electron gas, or jellium. In this case we have three 
degenerate phonon branches and due to the isotropy of the system the 
bare and BO normal modes are identical. The bare frequencies are 
given by the plasmon frequencies of the positive 
background charge~\cite{Mahan-book}:
\begin{align}
\w^{0}_{\n\blq}=\w^{0}=\sqrt{\frac{4\p Z^{2}_{\rm ion}n_{\rm 
ion}}{M_{\rm ion}}},
\end{align}
and are independent of $\blq$. The bare $e$-$ph$ coupling is 
given by~\cite{Mahan-book}
$g_{\n\blq}(\blk)=g_{q}=\sqrt{4\p}\;\w^{0}/q$,
and diverges for $q\to 0$. In jellium the phonon self-energy of 
Eq.~(\ref{phonself}) is diagonal with all equal entries 
$\P_{\blq}^{R}(\w)=|g_{q}|^{2}\tilde{\chi}^{R}_{00}(q,\w)$.
Evaluating $\tilde{\chi}^{R}_{00}(q,\w)$ in the random phase approximation,
\begin{align}
\tilde{\chi}^{R}_{00}(q\to 0,0)=-\frac{\a}{1+\a\frac{4\p}{q^{2}}},
\end{align}
with $\a=4p_{\rm F}/(2\p)^{2}$ and $p_{\rm F}$ the Fermi momentum,
one finds
\begin{align}
(\w_{\n \blq\to 0})^{2}&=(\w^{0})^{2}+\P^{R}_{q\to 0}(0)=(\w^{0})^{2}
\big[1+\frac{4\p}{q^{2}}\tilde{\chi}^{R}_{00}(q,0)\big]_{q\to 0}
\nn\\&=\lim_{q\to 0}(\w^{0}q)^{2}=0.
\end{align}
Thus, the bare optical phonon 
branch becomes an acoustic phonon branch. Accordingly, the unconventional 
screened coupling reads 
\begin{align}
\tilde{g}^{s}_{\n\bz}(\blk)=\lim_{q\to 0}
\left(\frac{\w_{\n\blq}}{\w^{0}}\right)^{2}
g_{q}=\lim_{q\to 0}q^{2}g_{q}=0.
\end{align}
Interestingly, in jellium the unconventional $\tilde{g}^{s}$ is the 
same as the conventional screened coupling 
$g^{s}=(1+v\tilde{\chi}^{R}_{00})g$.
The fpEE
correctly predicts that coherent 
acoustic phonons with zero momentum are not generated. 
We remark that the use of the model Ehrenfest equation, 
Eq.~(\ref{ehr1mod}), would have dramatic consequences in jellium 
since the bare $e$-$ph$ coupling diverges for $q\to 0$.

\section{Ehrenfest Equation for Polarons}
\label{polsec}

We consider a slightly doped crystal and relax the assumption of 
translational invariance. Due to doping,  the equilibrium 
value of the nuclear displacement $U_{i\a}$ of ion $i$ along 
direction $\a=x,y,z$ 
is, in general, nonvanishing. Its value can be determined from the 
fpEE  Eq.~(\ref{ehr1}), which in the basis $\{(i\a)\}$
reads~\cite{stefanucci_in-and-out_2023}
\begin{align}
\sum_{j\b}K^{0}_{i\a,j\b}U_{j\b}+\int 
d\blx g_{i\a}(\blr)\big(n^{\m}(\blx)-n^{0}(\blx)\big)=0.
\label{eqcond}
\end{align}
In this equation, $n^{\m}(\blx)$ is the 
equilibrium electronic density of the doped, and possibly distorted, system.
The label $\m$ stands for the chemical potential of the doped system;  
since $n^{0}$ is the equilibrium density of the undoped
system, notational consistency wants that we set the zero of energy 
at the chemical potential of the undoped system.
Equation~(\ref{eqcond}) can be manipulated using the BO Hessian $K$.
It is worth emphasizing that $\chi^{R}_{\rm 
clamp}$ in  Eq.~(\ref{omega}) is the response function 
at clamped nuclei of the {\em charged neutral} system.  Therefore
\begin{align}
\sum_{j\b}K_{i\a,j\b}U_{j\b}+\int 
d\blx g_{i\a}(\blr)\big(n^{\m}(\blx)-n^{0}(\blx)-\d 
n^{0}_{BO}(\blx,\blU)\big)=0,
\label{eqcond2}
\end{align}
where 
\begin{align}
\d n^{0}_{BO}(\blx,\blU)&\equiv\sum_{j\b}\int d\blx'\chi^{R}_{\rm clamp}(\blx,\blx';0)
g_{j\b}(\blr')U_{j\b}
\nn\\
&=n^{0}_{BO}(\blx,\blU)-n^{0}_{BO}(\blx),
\end{align}
with $n^{0}_{BO}(\blx,\blU)\equiv	
\bra\Q(\blR^{0}+\blU)|
\hat{n}(\blx)|\Q(\blR^{0}+\blU)\ket$ and 
$n^{0}_{BO}(\blx)=n^{0}_{BO}(\blx,\bz)$, see Eq.~(\ref{bodens}). 
Equation~(\ref{eqcond2}) is the second main result of our work.

So far no approximation has been made. We now assume that
the equilibrium 
BO density changes 
slightly due to the effects of phonons, and therefore
\begin{align}
n^{0}_{BO}(\blx)-n^{0}(\blx)\simeq 0.
\end{align}
Then, Eq.~(\ref{eqcond2}) reduces to
\begin{align}
\sum_{j\b}K_{i\a,j\b}U_{j\b}+\int 
d\blx g_{i\a}(\blr)\big(n^{\m}(\blx)-n^{0}_{BO}(\blx,\blU)
\big)=0.
\label{eqcond2app}
\end{align}
Equation~(\ref{eqcond2app}) 
must be coupled with equations for the 
electronic densities in such a way as to form a closed system of 
equations.  
In the quasiparticle approximation we can calculate the electronic densities 
from 
\begin{widetext}
\begin{subequations}
\begin{align}
	&\big[-\frac{\nabla^{2}}{2}+V+V_{\rm 
	H}[n^{\m}(\blx)]+\sum_{i\a}g_{i\a}U_{i\a}+\S(\e_{s})\big]\vf_{s}=\e_{s}\vf_{s}
	,\quad\quad n^{\m}(\blx)=\sum_{\e_{s}<\m}|\vf_{s}(\blx)|^{2},
	\\
	&\big[-\frac{\nabla^{2}}{2}+V+V_{\rm H}[n^{0}_{BO}(\blx,\blU)]
	+\sum_{i\a}g_{i\a}U_{i\a}+\S_{\rm 
	clamp}(\e^{BO}_{s})\big]\vf^{BO}_{s}=\e^{BO}_{s}\vf^{BO}_{s}
	,\quad\quad n^{0}_{BO}(\blx,\blU)=\sum_{\e^{BO}_{s}<0}|\vf^{BO}_{s}(\blx)|^{2}.
\end{align}
\label{polaroneq}
\end{subequations}
\end{widetext}
In these equations $V=V(\blr,\blR^{0})$ 
is the electron-nuclear potential of the equilibrium system, 
$V_{\rm H}[n]$ is the Hartree potential generated by a density $n$, 
$\S$ is the electronic exchange-correlation self-energy of 
the doped $e$-$ph$ system and $\S_{\rm clamp}$ is the electronic 
exchange-correlation self-energy of the charge neutral system at clamped nuclei. 
The nuclear displacements of the doped system are obtained from the 
self-consistent solution of Eqs.~(\ref{eqcond2app}) and 
(\ref{polaroneq}).

The inclusion of the Debye-Waller interaction can easily be dealt 
with at the mean-field level. It amounts to add the term
$\frac{1}{2}\sum_{i\a,j\b}g_{i\a;j\b}^{\rm DW}\bra 
\hat{U}_{i\a}\hat{U}_{j\b}\ket$ 
to the quasi-particle Hamiltonian. The calculation of $\bra 
\hat{U}_{i\a}\hat{U}_{j\b}\ket=U_{i\a}U_{j\b}+D_{i\a;j\b}(t,t^{+})$ necessitates 
to couple the above equations to the Dyson equation for the 
phononic Green's function $D$, unless we further approximate $\bra 
\hat{U}_{i\a}\hat{U}_{j\b}\ket\simeq U_{i\a}U_{j\b}$. Another interesting observation 
is that the Ehrenfest approximation, corresponding to set $\S=\S_{\rm 
clamp}=0$, 
already gives an indication of the tendency toward a distortion.

We conclude by comparing the fpEE polaron theory with the ab initio polaron 
theory initiated in Ref.~\cite{sio_ab-initio_2019} and further 
improved in 
Refs.~\cite{lafuente_unified_2022,lafuente_ab-initio_2022}. Consider 
a semiconductor in the ground state, with $s=v\blk$ ($s=c\blk$) the 
label of valence (conduction)
states with momentum $\blk$ of the charge neutral system. 
Under the assumption that a deformation of the 
lattice does not mix valence and conduction states, the 
eigenstates $\vf_{s}$ have the form
\begin{align}
\vf_{s}(\blx)=\sum_{\blk}\left\{\begin{array}{ll}
\sum_{c}A^{s}_{c\blk}\vf_{c\blk}(\blx) & \e_{s}>0 \\
\sum_{v}A^{s}_{v\blk}\vf_{v\blk}(\blx) & \e_{s}<0
\end{array}\right.,
\end{align}
where $\vf_{j\blk}(\blx)$, $j=c,v$, are the eigenstates of the 
charge neutral system. A similar expansion holds for the eigenstates 
$\vf^{BO}_{s}$. Accordingly, the densities of the doped and 
undoped systems read 
\begin{subequations}
\begin{align}
n^{\m}(\blx)&=\sum_{s}^{\e_{s}<0}\sum_{vv'\blk\blk'}A^{s\ast}_{v\blk}A^{s}_{v'\blk'}
\vf^{\ast}_{v\blk}(\blx)\vf_{v'\blk'}(\blx)
\nn\\&+
\sum_{s}^{0<\e_{s}<\m}\sum_{cc'\blk\blk'}A^{s\ast}_{c\blk}A^{s}_{c'\blk'}
\vf^{\ast}_{c\blk}(\blx)\vf_{c'\blk'}(\blx),\\
n^{0}_{BO}(\blx,\blU)&=
\sum_{s}^{\e_{s}<0}\sum_{vv'\blk\blk'}A^{BO,s\ast}_{v\blk}A^{BO,s}_{v'\blk'}
\vf^{BO\ast}_{v\blk}(\blx)\vf^{BO}_{v'\blk'}(\blx).
\end{align}
\end{subequations}
If the contribution from the valence bands is roughly the same for 
both densities, then Eq.~(\ref{eqcond2app}) simplifies to
\begin{align}
\sum_{j\b}K_{i\a,j\b}U_{j\b}&=-
\sum_{s}^{0<\e_{s}<\m}\sum_{cc'\blk\blk'}A^{s\ast}_{c\blk}A^{s}_{c'\blk'}
\nn\\
&\times \int 
d\blx\, g_{i\a}(\blr)\vf^{\ast}_{c\blk}(\blx)\vf_{c'\blk'}(\blx).
\label{eqcond3app}
\end{align}
Expanding the displacements in BO normal modes, see 
Eq.~(\ref{displdecnm}), and using the definitions in 
Eqs.~(\ref{Kprop}) and (\ref{proelphoncoup}), we can rewrite 
Eq.~(\ref{eqcond3app}) as 
\begin{align}
\w^{2}_{\n\blq}U_{\n\blq}=-\sum_{s}^{0<\e_{s}<\m}\sum_{cc'\blk}A^{s\ast}_{c\blk}
A^{s}_{c'\blk+\blq}g^{cc'}_{\n\blq}(\blk),
\end{align}
where $g^{cc'}_{\n\blq}(\blk)=\bra 
c\blk|g_{\n\blq}(\hat{\blr})|c'\blk+\blq\ket$ is the matrix element 
of the {\em bare} 
$e$-$ph$ coupling. 
This 
result agrees with Eq.~(40) of Ref.~\cite{lafuente_ab-initio_2022}.

\section{Summary and Conclusions}
\label{sumconcsec}

Although coherent phonons have been explored for over forty years, 
the first-principles equation of motion governing their formation and 
dynamics is derived and discussed in this work.
There are three main differences between the fpEE and the equation 
of motion based on model Hamiltonians. The first difference is the 
inclusion of nonadiabatic effects, leading to a renormalization 
of the BO frequencies and the introduction of a damping term in the 
quasi-phonon approximation. 
Both renormalization and damping are identical 
to those of quantum phonons. 
The 
second difference is that the interaction between coherent 
phonons and electrons 
is mediated by  an unconventional dynamically screened coupling. The 
crucial role of such screening has been made manifest in the 
homogeneous electron gas. The third difference is that only density 
variations beyond the linear order contribute to the   
electron-driven mechanisms of coherent phonon generation.
To second order in the external fields, two of these mechanisms are 
the impulsive stimulated Raman scattering and the displacive 
excitation~\cite{zeiger_theory_1992,kuznetsov_theory_1994,garrett_coherent_1996,merlin_generating_1997,ishioka2009coherent,zhai_coherent_2024}.

Another noteworthy feature of the fpEE is that the interaction 
between light and nuclei is mediated by the {\em dynamical} Born effective 
charge tensor. This quantity was  introduced only recently,  and 
in a completely different context -- specifically,
infrared optical conductivity and absorption 
spectra~\cite{bistoni_giant_2019,binci_first-principles_2021,wang_dynamical_2022,dreyer_nonadiabatic_2022}.

The fpEE has also been used to develop a theory of polaron 
distortions. Under reasonable assumptions our equations reduce to 
the ab initio approach of Ref.~\cite{lafuente_ab-initio_2022}. This 
agreement not only confirms the validity of the fpEE but also 
provides a unifying framework for studying light-induced and 
doping-induced coherent phonons.  

The reformulation of the fpEE presented in this work aims
to improve the understanding and modeling of coherent phonons in 
correlated systems from a first principles perspective. 
We should emphasize that the theory is based 
on the harmonic approximation and 
needs to be generalized to account for anharmonic 
effects. A first-principles treatment of anharmonicity, based 
on the expansion of the Hamiltonian for electrons and nuclei to third 
order in the fluctuation operators $U$ and $\D n$, 
as outlined in Ref.~\cite{stefanucci_in-and-out_2023}, is becoming 
increasingly urgent, given the growing interest in nonlinear 
phononics~\cite{forst_nonlinear_2011,subedi_theory_2014,juraschek_ultrafast_2017,vonhoegen_probing_2018,gu_nonlinear_2018,henstridge_nonlocal_2022}.

{\em Note added --} During the finalization of this manuscript we became
aware of the very recent work by Y. Pan, C. Emeis, S. 
Jauernik, M. Bauer and F. Caruso, arXiv:2502.01529 (2025), also deriving the 
first two terms of Eq.~(\ref{ehrqp}).

\acknowledgments

We acknowledge funding from Ministero Universit\`a e 
Ricerca PRIN under grant agreement No. 2022WZ8LME, 
from INFN through project TIME2QUEST, 
from European Research Council MSCA-ITN TIMES under grant agreement 101118915, 
and from Tor Vergata University through project TESLA.

\appendix

\begin{widetext}

\section{Derivation of Eq.~(\ref{ehr3})}
\label{fpEEapp}

Using Eq.~(\ref{chidyson}), we can rewrite Eq.~(\ref{dn1.1}) as 
\begin{align}
\D n^{(1)}(\blx t)&=\frac{1}{c}\sum_{\a}\int 
d(\blx't')\tilde{\chi}^{R}_{0\a}(\blx t,\blx't')
A^{\a}(t')
+\sum_{\bar{\n}\bar{\n}'}\int d(\bar{\blx}\bar{t})d(\bar{\blx}'\bar{t}')
\tilde{\chi}^{R}_{00}(\blx t,\bar{\blx}\bar{t})
g_{\bar{\n}\bz}(\bar{\blr})D^{0,R}_{\bar{\n}\bar{\n}'\bz}(\bar{t},\bar{t}')
g_{\bar{\n}'\bz}(\bar{\blr}')\D n^{(1)}(\bar{\blx}' \bar{t}'),
\label{dn1.2}
\end{align}
where in the second term we take into account that only the $\blq=0$ 
contribution survives due to the assumption of unbroken translational 
symmetry and the property
$g_{\n\blq}(\blr+\blR^{0}_{\bln})=e^{-i\blq\cdot\bln}
g_{\n\blq}(\blr)$.
Taking into account Eq.~(\ref{dn1.2}) we can then 
rewrite Eq.~(\ref{ehr1}) as
\begin{align}
\frac{d^{2}U_{\n\bz}(t)}{dt^{2}}
&=-\sum_{\n'}K^{0}_{\n\n'\bz}U_{\n'\bz}(t)
-\sum_{\bar{\n}\bar{\n}'}\int d\bar{t}d(\bar{\blx}'\bar{t}')
\P^{R}_{\n\bar{\n}\bz}(t,\bar{t})
D^{0,R}_{\bar{\n}\bar{\n}'\bz}(\bar{t},\bar{t}')
g_{\bar{\n}'\bz}(\bar{\blr}')\D n^{(1)}(\bar{\blx}' \bar{t}')
-\int d\blx \;g_{\n\bz}(\blr)
\D n^{(r)}(\blx t)+F_{\n}(t),
\label{ehr2}
\end{align}
with
\begin{align}
F_{\n}(t)&\equiv-\frac{1}{c}\sum_{\a}\int d\blx \;g_{\n\bz}(\blr)
\int d(\blx't')\tilde{\chi}^{R}_{0\a}(\blx t,\blx't')
A_{\a}(t')
-\frac{1}{c}\sum_{\a}Z_{\n\a}\frac{dA_{\a}(t)}{dt}.
\end{align}
Writing $\D n^{(1)}=\D n-\D n^{(r)}$, and  
observing that the solution of Eq.~(\ref{ehr1}) is~\cite{stefanucci_in-and-out_2023} 
\begin{align}
U_{\n\bz}(t)=\sum_{\bar{\n}'}\int d(\bar{\blx}'\bar{t}')
D^{0,R}_{\n\bar{\n}'\bz}(t,\bar{t}')
g_{\bar{\n}'\bz}(\bar{\blr}')\D n(\bar{\blx}' \bar{t}'),
\end{align}
we recover the first two terms of Eq.~(\ref{ehr3}). To prove that 
$F_{\n}(t)$ can be written as in Eq.~(\ref{scrforce}), we simply use the 
definitions in Eq.~(\ref{proelphoncoup}) for $g_{\n\blq}$ and 
Eq.~(\ref{bBOcharge}) for $Z_{\n\a}$.

\end{widetext}


%

\end{document}